\documentclass[epj]{svjour}
\usepackage{amsmath}
\usepackage{graphicx}
\usepackage{caption}
\usepackage{subcaption}
\usepackage[colorlinks=true, allcolors=blue]{hyperref}
\usepackage{lineno}
\usepackage[T1]{fontenc}
\usepackage{xspace}
\usepackage[utf8]{inputenc}
\usepackage{booktabs}
\usepackage{color}
\usepackage{euscript}
\usepackage{amsfonts}
\newcommand{\tomcca}{ToMCCA}
\newcommand{\dNde}{dN_\mathrm{ch}/d\eta}

\begin{document}
\title{ToMCCA: A Toy Monte Carlo Coalescence Afterburner}
\author{Maximilian Mahlein\inst{1,3} \and  Chiara Pinto\inst{1,2}\and Laura Fabbietti\inst{1}
%
}                     
\institute{Technical University of Munich, TUM School of Natural Sciences, Physics Department,\\ James-Franck-Stra{\ss}e 1, 85748 Garching b. M\"{u}nchen, Germany \and European Organisation for Nuclear Research (CERN), Geneva, Switzerland \and maximilian.horst@tum.de}
\date{Received: date / Revised version: date}
%
\abstract{
Antinuclei in our Galaxy may stem either from annihilation or decay of dark matter, or from collisions of cosmic rays with the interstellar medium, which constitute the background of indirect dark matter searches. 
Understanding the formation mechanism of (anti)nuclei is crucial for setting limits on their production in space. 
Coalescence models, which describe the formation of light nuclei from final-state interaction of nucleons, have been widely employed in high-energy collisions. 
In this work, we introduce ToMCCA (\textbf{To}y \textbf{M}onte \textbf{C}arlo \textbf{C}oalescence \textbf{A}fterburner), which allows for detailed studies of the nuclear formation processes without the overload of general-purpose event generators. 
ToMCCA contains parameterizations of the multiplicity dependence of the transverse momentum distributions of protons and of the baryon-emitting source size, extracted from ALICE measurements in pp collisions at $\sqrt{s} = 5 - 13$ TeV, as well as of the event multiplicity distributions, taken from the EPOS event generator. 
ToMCCA provides predictions of the deuteron transverse momentum distributions, with an agreement of $\sim5\%$ with the experimental data. The results of ToMCCA show that the coalescence mechanism in pp collisions depends only on the event multiplicity rather than on the collision system or its energy. This allows the model to be utilized for predictions at lower center-of-mass collision energies, which are the most relevant for the production of antinuclei from processes related to dark matter.
This model can also be extended to heavier nuclei as long as the target nucleus wave function and its Wigner function are known.
%
} 
\maketitle
\section{Introduction}
\label{intro}
The study and potential detection of antinuclei in cosmic rays has emerged to be an exciting field of research, offering a unique opportunity to explore possible signals from physics beyond the Standard Model. 
Antinuclei, consisting of antiprotons and antineutrons, are exceedingly rare in cosmic rays compared to their matter counterparts, and their presence in our Galaxy could be due to annihilation or decays of dark matter (DM) candidates~\cite{Donato:1999gy,Aramaki:2015pii}.
Indeed, the antinuclear flux expected from Standard Model processes due to collisions of high-energy cosmic rays (CRs) and the interstellar medium (ISM), is expected to be two to four orders of magnitude lower than the dark matter flux at low kinetic energies ($E_{\rm kin}<1$ GeV/$A$)~\cite{KorsmeierDeuteronFlux,Doetinchem_2020}. Both the CRs and the ISM are mostly made of protons \mbox{($\sim$ 90$\%$)} and $^4$He nuclei ($\sim$ 9$\%$), and only in a small percentage of heavier nuclei. Hence, the main contributing channels to the Standard Model background are collisions between protons and $^4$He nuclei (p--p, p--$^4$He, $^4$He--p, $^4$He--$^4$He), with center-of-mass energies of $\mathcal{O}(10)$ GeV~\cite{Serksnyte}.  
Understanding the formation mechanism of antinuclei is crucial for estimating the production cross-section from the Standard Model and Dark Matter contributions and for interpreting any future measurements of antinuclear flux from cosmic rays. Different experimental efforts are ongoing to measure the flux of antinuclei in cosmic rays, e.g. AMS-02~\cite{Kounine:2012ega}, GAPS~\cite{Hailey:2009fpa} and BESS-Polar~\cite{Abe:2011nx} experiments. However, antinuclei heavier than antiprotons have not been measured yet. 

Coalescence models have been widely used to study the formation of light nuclei starting from the production of individual nucleons in high-energy hadronic collisions~\cite{Coalescence1,Coalescence2,Coalescence3,Coalescence4,Coalescence5,Coalescence6,Coalescence7}. 
In these models, light nuclei are formed through the coalescence of nucleons that satisfy specific conditions, such as small relative momentum, spatial proximity, and spin-isospin alignment. In the simplest implementations of coalescence, only the momentum correlations are considered and the bound states are formed if the difference in momentum among the nucleons is below a given threshold, namely the coalescence momentum $p_0$. 
In the state-of-the-art implementations of the coalescence approach, the quantum-mechanical properties of the constituent baryons and those of the final bound states are taken into account and the coalescence probability is calculated from the overlap between the wave functions of individual (point-like) baryons and the Wigner density of the final-state cluster~\cite{Coalescence7,KachelriessLast,Blum:2019suo,Mrowczynski:2019yrr,Bellini:2020cbj,CoalescencePaper}. 
Monte Carlo (MC) simulations have proved to be powerful tools for investigating nuclear dynamics and particle production in high-energy hadronic collisions, as they offer a stochastic treatment of the complex dynamics in high-energy particle collisions while maintaining the full information on the space-time position and momenta of individual particles.
By combining the principles of MC event generators with coalescence models, recent descriptions of the momentum distributions of nuclei have drastically improved~\cite {KachelriessLast,CoalescencePaper}. 

In light of these advancements, we propose a new approach to MC-based coalescence, called ToMCCA (Toy Monte Carlo Coalescence Afterburner)\footnote{ToMCCA is available at https://github.com/HorstMa/ToMCCA-Public}.
ToMCCA is a Toy Monte Carlo that allows for a detailed study of the nuclear formation process without the overload of general-purpose event generators. 
In ToMCCA, the production of bound-states is implemented through a coalescence afterburner based on the Wigner function formalism. The event generation is customizable and the kinematics of the produced nucleons are tuned using parameterizations of (i) the multiplicity dependence of their transverse momentum ($p_{\rm T}$) distributions, (ii) their angular momentum correlations, (iii) the baryon-emitting source size, and (iv) the multiplicity distributions. 
Using such parameterizations and computing the probability of coalescence employing the Wigner function of the nucleus, ToMCCA provides predictions for the transverse momentum distributions of deuterons, for arbitrary multiplicity intervals covered by pp collisions. 
The parameterizations included in ToMCCA, based on measurements carried out with ALICE, make use of results from proton-proton (pp) collisions at a center-of-mass energy of \mbox{$\sqrt{s}=$ 5 TeV}~\cite{ALICE:2021ovi} and of those selected with the high multiplicity (HM) trigger in pp collisions at \mbox{$\sqrt{s}=$ 13 TeV}~\cite{nuclei_pp_13TeV_HM}. 
The performance of ToMCCA is tested against the deuteron $p_{\rm T}$ distributions of pp collisions at energies not included in the parameterizations, namely \mbox{$\sqrt{s}=$ 13 TeV}~\cite{Acharya:2020sfy} (with a minimum-bias trigger) and \mbox{$\sqrt{s}=$ 7 TeV}~\cite{7TeVDeuterons}. 
The ToMCCA model can be easily extended to heavier nuclei, as long as the nucleus wavefunction is known and its Wigner function is computed, and to lower center-of-mass collision energies, of the order of 10 GeV, that are the most relevant for the production of antinuclei from processes related to dark matter~\cite{Serksnyte}. 
\section{Working principle of ToMCCA}
\subsection{Coalescence model}
The coalescence afterburner model built in ToMCCA is based on the Wigner function formulation of coalescence. For more details, see Refs.~\cite{KachelriessLast,CoalescencePaper} and references therein. For convenience, we summarize here the basics of the coalescence model used in ToMCCA. 
The general idea is that the momentum distribution of deuterons in the final state ($d^3N/dP^3$) can be expressed by projecting the phase-space of the unbound proton-neutron (p-n) system ($W_{np}$) onto the Wigner function of the deuteron $D(\vec{q},\vec{r})$

\begin{equation}
\begin{split} 
    \label{eq:Yield}
    \frac{d^3N}{dP^3}=S\int d^3q\int d^3r_d \int d^3r D(\vec{q},\vec{r}) \times
    \\ \times W_{np}(\vec{P}_d/2+\vec{q},\vec{P}_d/2-\vec{q}, \vec{r}_d, \vec{r}).
\end{split}
\end{equation}
The S is the spin-isospin statistics factor, which equals 3/8 for deuterons, $\vec{P}_d$ is the total momentum, $\vec{q}$ is the relative momentum, $\vec{r}_d$ is the center of mass position, and $\vec{r}$ the distance of the p-n pair. The deuteron Wigner function depends only on the wavefunction of the deuteron for which there are multiple options, e.g., Gaussian, Hulthen~\cite{Heinz:1999rw}, Double Gaussian, Argonne $v_{18}$~\cite{PhysRevC.51.38}, etc. In this work, we use the Argonne $v_{18}$ wavefunction, which has been shown to have the best agreement with experimental data in Ref.~\cite{CoalescencePaper}. For the deuteron Wigner function, we use the parameterization reported in the appendix of Ref.~\cite{CoalescencePaper} for the Argonne $v_{18}$ wavefunction\footnote{The deuteron Wigner function for the Argonne $v_{18}$ wavefunction is available in the GitHub project at https://github.com/HorstMa/ToMCCA-Public.}.
We make the assumption that one can factorize the two-nucleon phase-space into its space and momentum terms. The p-n phase-space assumes the form

\begin{equation}
    \begin{split}   
    W_{np}(\vec{P}_d/2+\vec{q},\vec{P}_d/2-\vec{q}, \vec{r}_d, \vec{r})=&H_{np}(\vec{r},\vec{r}_d;r_0)\times
    \\\times G_{np}(\vec{P}_d/2+\vec{q},\vec{P}_d/2-\vec{q}).
    \end{split}
\end{equation} 
Folding the deuteron Wigner function with the nucleon phase-space distribution, it is possible to get the coalescence probability $\mathcal{P}(r_0,\vec{q})$ as a function of the source size $r_0$ and $\vec{q}$

\begin{equation}
    \label{eq:CoalescenceProbability}
    \mathcal{P}(\vec{q},r_0)=\int d^3q\int d^3r \int d^3r_d D(\vec{q},\vec{r}) H_{np}(\vec{r},\vec{r}_d;r_0).
\end{equation}
The source size is described in Sec.~\ref{sec:Source}. 
Such probability can then be applied on any p-n pair, on a pair-by-pair basis, to predict deuteron transverse momentum distributions.

\subsection{The ToMCCA toy Monte Carlo}
\label{sec:tomcca}
The use of a toy MC instead of traditional MC event generators, such as Pythia~\cite{Skands:2014pea,Sjostrand:2014zea} or EPOS~\cite{EPOS}, allows for a much faster computation, as only processes strictly necessary for coalescence are computed. 
Furthermore, such a lightweight generator can be modified and customized easily by the end-user, allowing for much more differential studies.  
In this work, we developed a Toy Monte Carlo that reproduces the deuteron momentum distributions employing parameterizations of the following parameters:
\begin{itemize}
    \item Charged-particle multiplicity of events,
    \item Nucleon momentum distributions,
    \item Angular correlations among nucleons,
    \item Hadron-emitting source size,
    \item Nuclear wavefunction. 
\end{itemize}
The parameterizations included in ToMCCA are based on experimental measurements from the ALICE collaboration, except for the parameterization of the charged-particle multiplicity distributions, which is based on EPOS 3, as described in detail in Sec.~\ref{sec:Kinematics}. 

\subsubsection{Event multiplicity}
\label{sec:Multiplicity}
The starting point of ToMCCA is the determination of the charged-particle multiplicity of the event that is generated. This number is usually given by the number of charged particles per unit of pseudorapidity $\eta$, e.g., measured in a range of $|\eta|<0.5$ ($(dN_{ch}/d\eta)_{|\eta|<0.5}$). 
Particle spectra are often measured in a different kinematic range, such as $|y|<0.5$, where $y$ is the rapidity relative to the beam axis.
To transform from $|\eta|<0.5$ to $|y|<0.5$, a scaling constant $\alpha$ has to be determined such that $\alpha\cdot(dN_{ch}/d\eta)_{|\eta|<0.5} = (dN_{ch}/dy)_{|y|<0.5}$~\cite{MultiplicityTransformationPaper}. This can be computed using event generators. For the case of ToMCCA, a value of $\alpha=1.192$ is determined using EPOS 3~\cite{EPOS}. 

There are multiple ways to reproduce each given $dN_\mathrm{ch}/d\eta$ value. Two simple approaches are included in ToMCCA: (i) fixed multiplicity, i.e., generating exactly $\alpha\dNde$ particles per event, and (ii) Poissonian multiplicity, i.e., generating a Poissionian distribution around the chosen multiplicity value. However, since coalescence is a non-linear process, i.e., it depends on the number of pairs rather than single particles, the overall shape of the distribution plays a significant role. Thus, a more realistic approach is needed. An improvement can be reached by using an event generator, such as EPOS 3. Using an unbiased triggering method, such as the one employed by the ALICE experiment~\cite{nuclei_pp_13TeV_HM}, the chosen mean charged-particle multiplicities can be triggered outside of the nucleus measurement region ($-3.7<\eta<-1.7$ and $2.8<\eta<5.1)$. The shapes of the resulting midrapidity distributions can be well described using the Erlang distribution~\cite{ErlangMultiplicity}, given by

\begin{equation}
    f(x;k,\lambda)=\frac{\lambda^k x^{k-1} e^{-\lambda x}}{(k-1)!}.
\end{equation}
While technically the Erlang distribution is only defined for $k\in\mathbb{N}^+$, it can be extended to $k\in\mathbb{R}_{\geq0}$ without introducing any poles, as long as $x>0$. This further requires to generalize the factorial $(k-1)!$ to the gamma function $\Gamma(k)$, valid for $k\in\mathbb{R}_{\geq0}$. With these modifications, the multiplicity dependence of the parameters of the Erlang function $k$ and $\lambda$ can be modeled with a power law function (given by Eq.~\ref{eq:PowerLaw}). The parameters of the power law function can be found in Tab.~\ref{tab:ErlangParameters}. 

\begin{table}[h]
    \centering
    \begin{tabular}{lll}
    \noalign{\smallskip}\hline
         Parameter& k & $\lambda$  \\
         \noalign{\smallskip}\hline\noalign{\smallskip}
         A&0&0\\
         B& 0.77544 & 0.72653\\
         C& 0.74669& -0.23343\\
         \noalign{\smallskip}\hline
    \end{tabular}
    \caption{Parameterization of the Erlang distribution using power law functions.}
    \label{tab:ErlangParameters}
\end{table}

\subsubsection{The ToMCCA event}
\begin{figure*}
    \centering
    \includegraphics[width=1\textwidth]{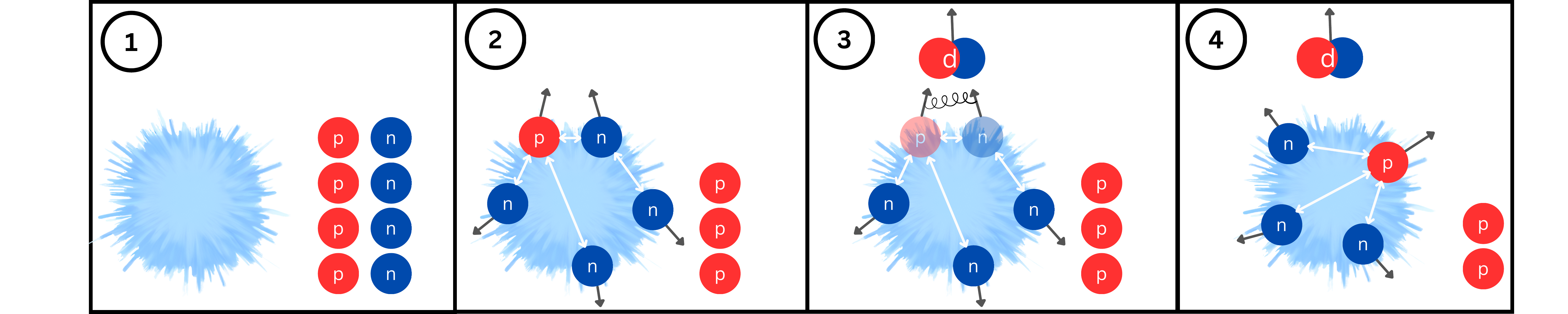}
    \caption{1) Starting situation after the number of protons (red) and neutrons (blue) have been determined. 2) For each proton, a new set of neutrons is generated. 3) One proton-neutron pair that satisfies the coalescence condition forms a deuteron. 4) Next proton of the event gets a new set of neutrons. The amount of neutrons has been reduced by one since one neutron was bound to a proton to form the deuteron.}
    \label{fig:EventLoop}
\end{figure*}

A schematic view of the event loop in ToMCCA is shown in Fig.~\ref{fig:EventLoop}. First, the number of nucleons (protons and neutrons separately) in a given event is calculated starting from the particle multiplicity of the event (Sec.~\ref{sec:particleproduction}). In panel (1) of Fig.~\ref{fig:EventLoop}, an exemplary situation in which four protons (red) and four neutrons (blue) are produced is depicted. Then, the 3-dimensional momentum of the first proton and all neutrons is determined (Sec.~\ref{sec:Kinematics}), as well as their relative distances (Sec.~\ref{sec:Source}), as sketched in panel (2) of Fig.~\ref{fig:EventLoop}. Using these momenta ($q$) and distances ($r_0$), the coalescence probability $\mathcal{P(\it{q}, r_0)}$ is calculated for each pair using Eq.~\ref{eq:CoalescenceProbability}. A random number $r\in[0,1]$ is drawn, and if $r<\mathcal{P(\it{q}, r_0)}$ the pair is accepted as a deuteron. Otherwise, the proton gets checked against other neutrons. If multiple neutrons satisfy the condition of coalescence with the same proton, only the first pair forms the deuteron. However, this is a very rare occurrence comparable to the $^3$He and $^3$H production rates ($\sim1/10^{-7}$events~\cite{nuclei_pp_13TeV_HM}). After all p-n pairs have been combined, a new proton and a completely new set of neutrons are created, and the same procedure is repeated. If a deuteron is created in the first loop, the number of neutrons available for the following protons in this event is reduced by one. In panel (3) of Fig.~\ref{fig:EventLoop}, a sketch of a deuteron being formed is shown. The subsequent event in Fig.~\ref{fig:EventLoop} (4) shows the second proton with now only three, and completely new, neutrons.

\subsubsection{Particle production}
\label{sec:particleproduction}
In ToMCCA, the exact composition of charged particles is not important. Only the final number of protons and neutrons that are generated is of interest. 
There are four methods implemented in ToMCCA to determine the number of nucleons: uncorrelated emission, string fragmentation, quark recombination, and tuned emission. The methods used for particle production are described in detail below. 

\textbf{1. Uncorrelated Emission} 
\\In this method, for every charged particle created, there is a probability for it to be a proton given by 

\begin{equation}
p_{prot}=(dN/dy)/(dN_{ch}/dy),
\label{eq:p_prot}
\end{equation}
where d$N$/d$y$ is the production \textit{yield} of protons, anchored to the measured production spectra, as will be discussed in Sec.~\ref{sec:Kinematics}. 
A mirroring technique is used to determine the number of neutrons, where $N_{ch}$ neutral particles are created and assigned to be neutrons in the same manner as protons. Notably this method creates no correlation between the event-by-event number of protons and that of neutrons, and therefore between protons and deuterons. The impact of this could be quantified by studying the deuteron number fluctuations and correlations between nucleons and deuterons on an event-by-event basis, as it has been done in Ref.~\cite{ALICE:2022amd} for Pb--Pb collisions. 

\begin{figure}[!hbt]
    \centering
    \includegraphics[width=0.45\textwidth]{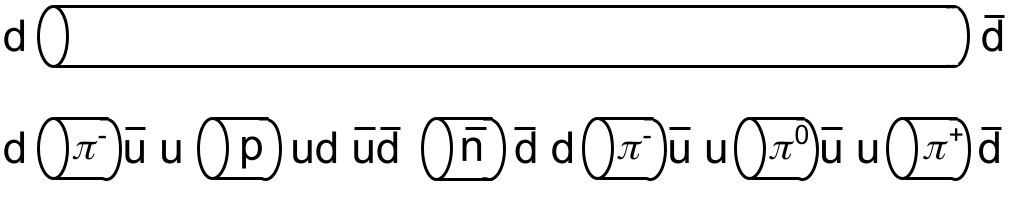}   
    \caption{An example of the fragmentation of a $\texttt{d}\Bar{\texttt{d}}$ string. Mesons are indicated for clarity but are ignored for coalescence in ToMCCA.} 
    \label{fig:StringFragmentation}
\end{figure}

\textbf{2. String Fragmentation}
\\The second method implemented in ToMCCA is called String Fragmentation. The method mimics much more complex hadronization methods employed by general- purpose event generators, such as Pythia or EPOS. Here, one large string is created, and either a \texttt{u}($\Bar{ \texttt{u}}$) or a  \texttt{d}($\Bar{ \texttt{d}}$) quark is placed at one end, and its antimatter counterpart at the other end. This string is fragmented in a given number of locations, which equals the total number of particles created $dN_{\mathrm{ch}}/dy$. On each fragmentation location, a quark-antiquark ($ \texttt{q}\Bar{ \texttt{q}}$) or diquark-antidiquark ($ \texttt{dq}\overline{ \texttt{dq}}$) pair is created. The order of the pairs is determined so that a color-neutral particle is created at all times, i.e., diquarks are paired with quarks, and single quarks are paired with single antiquarks. To avoid tetraquark configurations, no diquark can be created directly after another diquark. After all the fragmentations, the triplets created from diquarks are assigned to the corresponding particles depending on the quark content. The combinations \texttt{ddd} and \texttt{uuu} are assigned to neutrons and protons respectively, since these are $\Delta^-$ and $\Delta^{++}$ resonances, which decay exclusively into the assigned nucleons. Mesons that would be created with this method are ignored for simplicity. An example of a string with an initial $\texttt{d}\Bar{\texttt{d}}$ configuration is shown in Fig.~\ref{fig:StringFragmentation}. 

\noindent This method determines the total proton yield starting from the probability of creating a diquark. Due to edge effects in the fragmentation algorithm, baryon production at low multiplicities gets suppressed. The diquark probability needs to be modified and parameterized as a function of multiplicity, in order to recover the production yield measured at low multiplicity. This probability is parameterized such that the proton yields as a function of multiplicity correspond to the one from the uncorrelated emission method through the equation:

\begin{equation}
\begin{split}
    p_\mathrm{diquark}=p_\mathrm{prot,modified}(dN/dy) =
    \\= p_\mathrm{prot}(dN/dy) \cdot (A+B\ (dN/d\eta)^{C}),
\end{split}
    \label{eq:ProbabilityParameterization}
\end{equation}
where $p_{\mathrm {prot}}$ is given by Eq.~\ref{eq:p_prot}.
The obtained parameters can be found in Table~\ref{tab:FitHadMethod}.

\textbf{3. Quark recombination}
\\The Quark Recombination method is based on the idea that a sea of \texttt{u} and \texttt{d} quarks is created and recombined randomly to create hadrons. The total number of each quark species is $N_d=N_u=dN_\mathrm{ch}/dy$, where $dN_\mathrm{ch}/dy$ is the number of charged particles as described above. First, a \texttt{u} and a \texttt{d} quark are combined randomly, and then, with probability $p_\mathrm{prot, modified}$ given by Eq.~\ref{eq:ProbabilityParameterization}, a third \texttt{u}(\texttt{d}) quark is added to the \texttt{ud} pair, creating a proton (neutron). This is repeated until no \texttt{u}\texttt{d} pairs are left.
The fit parameters of Eq.~\ref{eq:ProbabilityParameterization}, for the parameterization of the probability of adding one quark to the \texttt{ud} pair, are also reported in Tab.~\ref{tab:FitHadMethod}.

\begin{table}[!hbt]
    \centering
    \begin{tabular}{lll}
    \hline\noalign{\smallskip}
         Parameter &String Fragmentation  & Quark recombination\\
         \noalign{\smallskip}\hline\noalign{\smallskip}
         A& 2.13957998 &0.87627309\\
         B& -3.8557517 & 1.01566945\\
         C& -2.0928103 & -0.54537776\\
         \noalign{\smallskip}\hline
    \end{tabular}
\caption{Fit parameters of Eq.~\ref{eq:ProbabilityParameterization}, for the parameterization of the diquark probability (string fragmentation method) and of the probability of adding a quark (quark recombination method).}
\label{tab:FitHadMethod}
\end{table}
\textbf{4. Tuned Emission}
\\Both the String Fragmentation and Quark Recombination methods by construction have a fixed Pearson correlation coefficient $\rho_{pn}$ (Eq.~\ref{eq:Pearson}), which is a measure of the correlation between the number of protons and neutrons in a given event. A negative coefficient means that the presence of a proton suppresses the creation of new neutrons. To get around this feature, a fourth method is developed, namely the tuned emission, which contains a free parameter $a$ to adjust the level of nucleon suppression. It is based on the uncorrelated emission, but the probability for a neutron $p_{neut}$ is modified depending on the number of protons in the event via \mbox{$p_{neut,tuned}=(a(N_{p}-dN/dy)+1)p_{neut}$}. Since the String Fragmentation and Quark Recombination models show an unreasonably strong suppression of nucleon pairs at low multiplicities, the value $a$ was set in such a way that its Pearson coefficient lies roughly in the middle between String Fragmentation and Uncorrelated Emission.
The exact parameterization is $a=1.508 \, (dN_{ch}/dy)^{-1}$. The difference in the final deuteron yields between all these methods (1 to 4) ranges from $<10\%$ for $dN_{ch}/d\eta|_{|\eta|<0.5}>30$ to $300\%$ for $dN_{ch}/d\eta|_{|\eta|<0.5}\sim3$. This shows that the exact degree of suppression is a valuable input to this model. 
The parameter $a$ can be tuned to a measurement of the Pearson coefficient of proton and deuteron $\rho_{\rm{pd}}$ as the correlation between proton and neutron is transferred into the correlation between proton and deuteron. Such measurements have been performed in the past on Pb--Pb data~\cite{PbPbCorrelations}, but not yet in pp collisions.
\\The Pearson correlation coefficient can be calculated as

\begin{equation}
    \rho_{\rm{pn}}=\langle(n_p-\langle n_p\rangle)(n_n-\langle n_n\rangle)\rangle/\sqrt{\kappa_{2p}\kappa_{2n}},
    \label{eq:Pearson}
\end{equation}
where $\kappa_{2a}$ is the second order cumulant of the particle multiplicity distribution $\kappa_{2a}=\langle(n-\langle n\rangle)^2\rangle$.
The calculated proton-neutron Pearson coefficients $\rho_{\rm{pn}}$ for each method can be found in Tab.~\ref{tab:PearsonCoefficients}. 
\\In the following, all results are obtained using the Tuned Emission method. 

\begin{table}[h]
    \centering
\begin{tabular}{lllll}
\noalign{\smallskip}\hline
     &Unc. Em.&\hspace*{-0.1cm}String Fragm.&\hspace*{-0.1cm}Quark Rec.&\hspace*{-0.1cm}Tuned Em.\\
     \noalign{\smallskip}\hline\noalign{\smallskip}
     $\rho_{\rm{pn}}$&0&\hspace*{-0.1cm}-0.052&\hspace*{-0.1cm}-0.058&\hspace*{-0.1cm}-0.024\\
     \noalign{\smallskip}\hline
\end{tabular}
    \caption{The Pearson correlation coefficient between protons and neutrons for different hadronization methods.}
    \label{tab:PearsonCoefficients}
\end{table}


\subsubsection{Kinematics}
\label{sec:Kinematics}
Once the number of protons and neutrons is calculated, their kinematics have to be determined. In the following, the procedure used to assign the three-dimensional momentum to nucleons and the angular correlations to the nucleon pairs is described. 

\textbf{1. Nucleon momentum distributions}
\\The $p_\mathrm{T}$ distributions of nucleons are described using a Levy-Tsallis function (Eq.~\ref{eq:LevyTsallis}). However, the shape of the $p_\mathrm{T}$ distribution evolves with the multiplicity of the event~\cite{ALICE:2018pal}. Thus, ToMCCA requires a parameterization of the shape of the transverse momentum distributions of nucleons as a function of multiplicity. For this, the proton spectra measured by ALICE in pp collisions at a center-of-mass energy of \mbox{$\sqrt{s}=$ 5 TeV}~\cite{ALICE:2021ovi} and those selected with the high multiplicity (HM) trigger in pp collisions at \mbox{$\sqrt{s}=$ 13 TeV}~\cite{nuclei_pp_13TeV_HM} have been fitted using Eq.~\ref{eq:LevyTsallis}. The evolution of the Levy-Tsallis parameters (d$N$/d$y$, $n$, $T$) has been modeled as a function of $dN/d\eta$ according to a power law function (Eq.~\ref{eq:PowerLaw}).
The values of the parameters can be found in Tab.~\ref{tab:YieldParameters}. 
Due to a high correlation between the parameters \textit{T} and \textit{n}, the latter was constrained to Eq.~\ref{eq:PowerLaw}, only letting the parameter \textit{A} vary slightly ($7.3\pm0.1$) and fixing \textit{C}=$-$1. This is to ensure that the \textit{T} parameter evolves smoothly as a function of multiplicity. Furthermore, depending on the hadronization model used for the particle production (Sec.~\ref{sec:particleproduction}), the parameter \textit{T} needs to be modified since the models can show edge effects that alter the yield, especially at low multiplicities. For this, \textit{T} gets modified as

\begin{equation}
    T_\mathrm{mod}(dN/d\eta)=T(dN/d\eta) \cdot (A+B(dN/d\eta)^{C}).
    \label{eq:T}
\end{equation}

\noindent The resulting parameterization can be found in Tab.~\ref{tab:YieldParameters}. Since in the modification term $A\approx1$ and \mbox{C $<$ 0}, this correction only modifies low multiplicity events, while the modification term tends to 1 for high multiplicity events.

\begin{table}[h]
    \centering
    \begin{tabular}{llllll}
     \noalign{\smallskip}\hline
    & dN/dy&n&T&$T_\mathrm{mod}^\mathrm{QuarkReco}$&$T_\mathrm{mod}^\mathrm{StringFrag}$\\
     \noalign{\smallskip}\hline\noalign{\smallskip}     
     A&  --&7.3230690&--&0.9953433&0.9947258\\
     B&  0.06152&2.0855418&0.10805&-0.7380489&-4.341114\\
     C&  0.95621&-1&0.36889&-2.0069334&-4.555843\\
      \noalign{\smallskip}\hline
     
    \end{tabular}
    \caption{Values of the parameters for the yield parameterization. The parameters $T_\mathrm{mod}^\mathrm{QuarkReco}$ and $T_\mathrm{mod}^\mathrm{StringFrag}$ refer to the modifications of the temperature parameter $T$ of Eq.~\ref{eq:T}, for the quark recombination and string fragmentation hadronization methods, respectively. }
\label{tab:YieldParameters}
\end{table}



\noindent In addition to the transverse momentum, the rapidity $y$ is drawn from a flat distribution $y\in[-0.5,0.5]$. This is well motivated for measurements at midrapidity~\cite{STARRapidity}. 

\textbf{2. Angular Correlations}
\\To eventually fix the full 3-dimensional momentum information, the relative angle between the proton and the neutron needs to be determined. Experimentally, this is accessed using the angular correlation function~\cite{13TeVAngularCorrelations}

\begin{equation}
    C(\Delta\varphi)= \frac{S(\Delta\varphi)}{B(\Delta\varphi)}, 
\end{equation}
where $S(\Delta\varphi)$ is the same event $\Delta\varphi$ distribution, i.e., the angular distribution of the azimuthal angle between two particles in the same event. Conversely, $B(\Delta\varphi)$ is the $\Delta\varphi$ distribution of pairs from mixed events, i.e., relative angles between particles from different events. The latter removes trivial correlations from the same event distribution and reduces them to the genuine correlation. However, for a model like \tomcca, the actual angle between particles in the same event is needed. Hence only the same event angular distribution $S(\Delta\varphi)$ is required. In order to isolate this distribution, the measured $C(\Delta\varphi)$ distribution is multiplied by a $B(\Delta\varphi)$ obtained from event generators. This method is applicable since $B(\Delta\varphi)$ contains only trivial correlations due to collision geometry and energy conservation that are well reproduced in event generators. For \tomcca, $B(\Delta\varphi)$ is obtained again from EPOS 3 and assumed to be independent of the event multiplicity. The $C(\Delta\varphi)$ for $pp+\overline{p}\overline{p}$ pairs is measured by the ALICE collaboration~\cite{13TeVAngularCorrelations}, as a function of multiplicity. However, since measurements in all multiplicity classes are compatible with each other within their uncertainties, the data measured in the 20-40\% multiplicity class is used for all multiplicities.
The final same event distribution $S(\Delta\varphi)=C(\Delta\varphi)\times B(\Delta\varphi)$ is expressed as

\begin{equation}
    \begin{split}  
    S(\Delta\varphi)= \underbrace{(N_0 \sin(\Delta\varphi - \pi/2) + 1)}_{C(\Delta\varphi)}\times\\
    \times     \underbrace{(N_1|\Delta\varphi|+B)(N_2 e^{-\Delta\varphi^2/A^2})}_{B(\Delta\varphi)}.
    \end{split}
\end{equation}
The parameters can be found in Tab.~\ref{tab:AngularCorrelationParameters}.

\begin{table}[h!]
    \centering
    \begin{tabular}{ll}
        \noalign{\smallskip}\hline
        Parameter & $B(\Delta\varphi)$ \\
        \noalign{\smallskip}\hline\noalign{\smallskip}
        A & 0.85747601 \\
        B & 5.7197689 \\
        $N_1$ & -0.89304215 \\
        $N_2$ & 0.26041062 \\
        $N_0$ & 0.17701025\\
        \noalign{\smallskip}\hline
    \end{tabular}
    \caption{Values for the parameterization of the angular correlation. }
    \label{tab:AngularCorrelationParameters}
\end{table}

\section{Source size}
\label{sec:Source}
The last key ingredient for coalescence is the source size $r_0$, which is a measure of the distance between the particles of the pair. In ToMCCA, the source is assumed to be distributed according to a Gaussian function, and thus the distances $r$ between particles follow the distribution~\cite{sourceSizeHMpp}

\begin{equation}
    S(r;r_0)=\frac{4\pi r^2}{(4\pi r_0^2)^{3/2}}e^{-\frac{r^2}{4r_0^2}},
\end{equation}
from which distances are drawn at random. The source size in small systems has been measured as a function of the average transverse mass\footnote{The average transverse mass is defined as $\langle m_\mathrm{T}\rangle = \sqrt{\left(\frac{p_\mathrm{T,1}+p_\mathrm{T,2}}{2}\right)^2+\left(\frac{m_1+m_2}{2}\right)^2}$ of the pair~\cite{sourceSizeHMpp}. In the case of deuteron coalescence, the masses of the constituent nucleons are almost equal, $m_1=m_2=m_N\approx0.938~$GeV$/c^2$, and the transverse momentum of the deuteron is given by the sum of those of the constituents, \mbox{$p_\mathrm{T,d}=p_\mathrm{T,1}+p_\mathrm{T,2}$}} in pp collisions at $\sqrt{s}~=~13$ TeV with a high multiplicity trigger~\cite{sourceSizeHMpp}. These collision events have a mean charged-particle multiplicity of about $dN_{ch}/d\eta\sim30$. On the other hand, minimum bias pp collisions have a mean multiplicity of $dN_{ch}/d\eta\sim7$ and it is expected that the source size depends on multiplicity~\cite{Bellini:2018epz}. Thus, a description of the source size for arbitrary multiplicity is needed to accurately describe the deuteron production. For this purpose, the coalescence model is reverted, leaving the source size as a free parameter and including the deuteron momentum distributions in the fitting procedure of \tomcca. The deuteron spectra are taken from the ALICE measurements in pp collisions at \mbox{$\sqrt{s}=$ 5 TeV}~\cite{ALICE:2021ovi} and from the HM sample of pp collisions at \mbox{$\sqrt{s}=$ 13 TeV}~\cite{nuclei_pp_13TeV_HM}, spanning a range of $dN_\mathrm{ch}/d\eta$ from 2.4 to 35.8~\cite{nuclei_pp_13TeV_HM,Acharya:2020sfy}. 

\begin{figure}[h]
    \centering
    \includegraphics[width=0.45\textwidth]{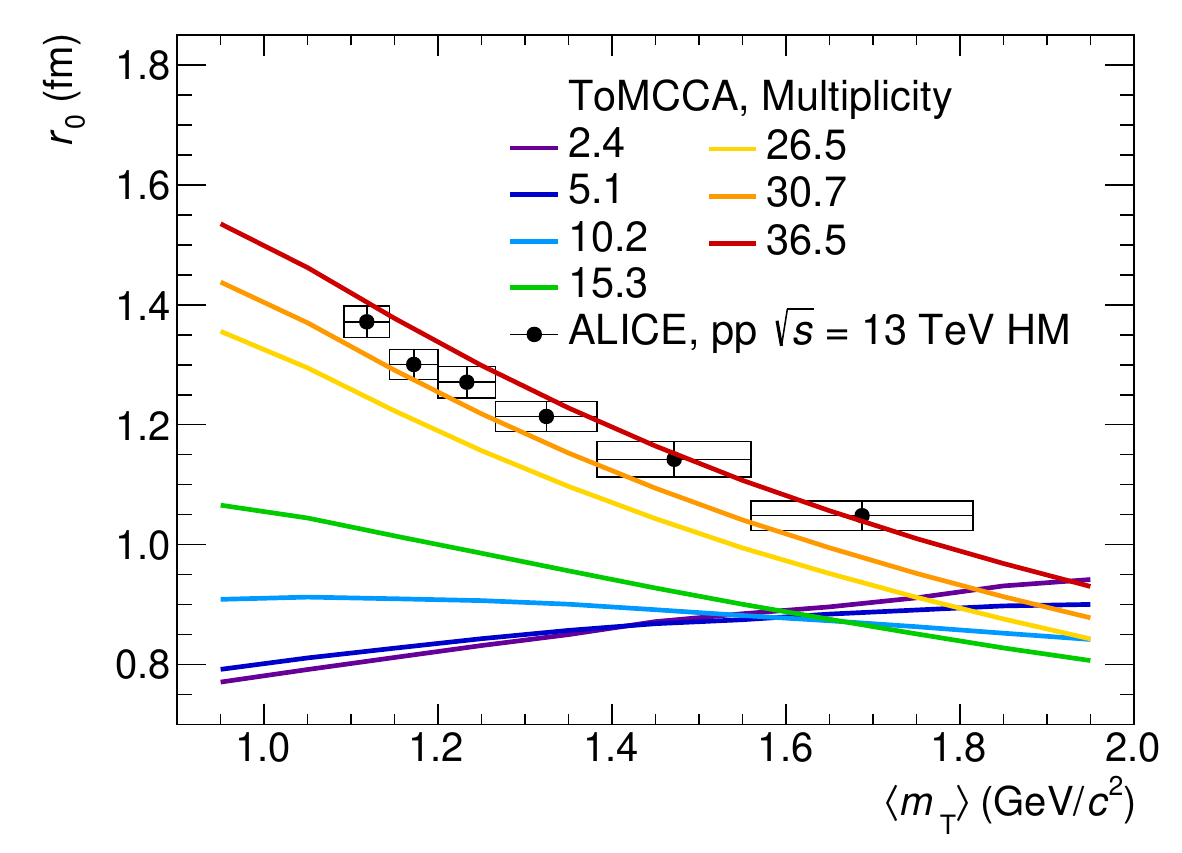}
    \caption{Scaling of the baryon-emitting source size as a function of $\langle m_T\rangle$, for arbitrary multiplicity values. The predictions of ToMCCA are compared to the measured source size corresponding to an event multiplicity of about 30~\cite{sourceSizeHMpp}. }
    \label{fig:SourceSizeScaling}
\end{figure}

The fitting procedure consists of several steps, as described in the following.
First, for a given multiplicity, the deuteron $p_\mathrm{T}$ distributions are generated for source sizes from $r_0$= 0 to 1.8 fm.
Then, for each value of $p_\mathrm{T}$, the $\chi^2=\frac{Y_\mathrm{ToMCCA}-Y_\mathrm{ALICE}}{\Delta Y_\mathrm{ALICE}}$ (being $Y$ the yield for a given $p_\mathrm{T}$, either in ToMCCA or in data, and $\Delta Y$ the uncertainty of the data points) is calculated for all tested source sizes. The resulting distribution of $\chi^2$ as a function of the source size, $\chi^2(r_0)$, can be fitted with a polynomial function of second order, $y=a(x-c)^2$. The parameter $c$ gives the best fitting source size and $1/\sqrt{a}$ corresponds to the $1\sigma$ deviation. This procedure is repeated for every value of $p_\mathrm{T}$ covered by the measurement. The resulting source size scaling as a function of $\langle m_\mathrm{T} \rangle$ is fitted using a power law (Eq.~\ref{eq:PowerLaw}, with parameter A=0).
Finally, the procedure is repeated for all multiplicities in the datasets of pp collisions at $\sqrt{s}=$ 5 TeV and HM at $\sqrt{s}=$ 13 TeV. The resulting power law parameters $B$ and $C$ are fitted as a function of $dN/d\eta$ with Eq.~\ref{eq:Root3Sigmoid} and Eq.~\ref{eq:Sigmoid}, respectively. The functional form of Eq.~\ref{eq:Root3Sigmoid} has been chosen to reproduce the observation of $r_0\propto N_{ch}^{1/3}$ at large multiplicities~\cite{HydrodynamicsSource} and its saturation to a minimal source size $r_0\sim R_p=0.87$ fm~\cite{ProtonRadius} similar to the proton radius ($R_p$). The parameters are reported in Tab.~\ref{tab:SourceParameter}. With such parameterization, it is possible to obtain the $\langle m_T\rangle$ scaling of the source size for arbitrary multiplicity values, as shown as an example in Fig.~\ref{fig:SourceSizeScaling}. The predictions of ToMCCA are compared to the source size measured by ALICE~\cite{sourceSizeHMpp} in the HM data sample of pp collisions, corresponding to an average multiplicity of about 30 charged particles, showing very good agreement with the results of the model. To date, there are no experimental measurements of the $\langle m_T\rangle$-scaling of the source radii as a function of the average charged-particle multiplicity in minimum bias pp collisions at the LHC. Future measurements of this quantity with the LHC Run 3 data will be useful to make a quantitative comparison to the predictions of ToMCCA. 

The uncertainties associated with these predictions stem from the uncertainty of the measurement of the proton and deuteron production yields. The model uncertainties are estimated by repeating the fitting procedure using a parameterization of the lower and upper bounds of the $dN/dy$ parameter in Eq.~\ref{eq:LevyTsallis} and taking the uncertainty from the $\chi^2$ minimization described above. The obtained global uncertainties are 3.5\% from the protons and 4.4\% from the deuterons, which, added in quadrature, give \\\mbox{$\delta B(dN/d\eta)=5.7\%$}. The C parameter is unchanged within these variations. Propagating this uncertainty to the predictions of the deuteron spectra, by varying $B(dN/d\eta)$ by 5.7\%, changes the deuteron yield by $\pm 4.6\%$.

\begin{table}[h!]
    \centering
    \begin{tabular}{lll}
    \noalign{\smallskip}\hline
         Parameter& B($dN/d\eta$) & C($dN/d\eta$)  \\
         \noalign{\smallskip}\hline\noalign{\smallskip}
         A& 0.59781 &-0.85246\\
         B& 0.58983 &0.83824\\
         C& 1.26481 & 4.25301\\
         D& -0.48262 &0.182242\\
         N& -- &0.99022\\
         M & 10.8889 & --\\
         \noalign{\smallskip}\hline
    \end{tabular}
    \caption{Parameterization as a function of multiplicity of the power law parameters describing the source size $r_0$.}
    \label{tab:SourceParameter}
\end{table}

\section{Results}
\subsection{Transverse momentum distributions and coalescence parameter}
\label{results}
\begin{figure}[h]
    \centering
    \includegraphics[width=0.5\textwidth]{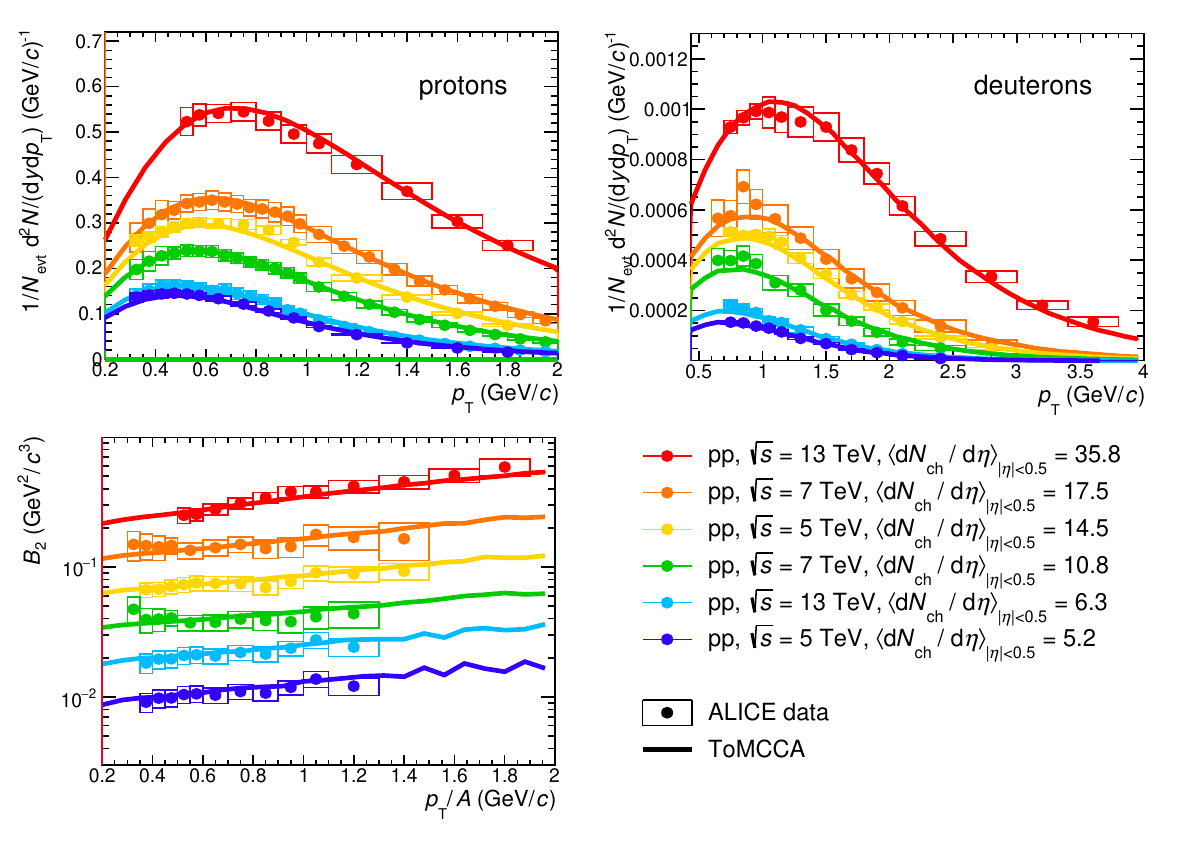}
    \caption{Results of ToMCCA in different multiplicity classes, for the transverse momentum spectra of protons (top left) and deuterons (top right), and the coalescence parameter $B_2$ (bottom), compared to data measured by ALICE. The values of the average charged-particle multiplicity corresponding to each multiplicity class in the three different collision energies are indicated in the legend. The values of $B_2$ are scaled for better visibility, by the following factors, going from the top to the bottom: 42, 16, 8, 4, 2, 1. }
    \label{fig:results_allclasses}
\end{figure}

\begin{figure}[h]
    \centering
    \includegraphics[width=0.5\textwidth]{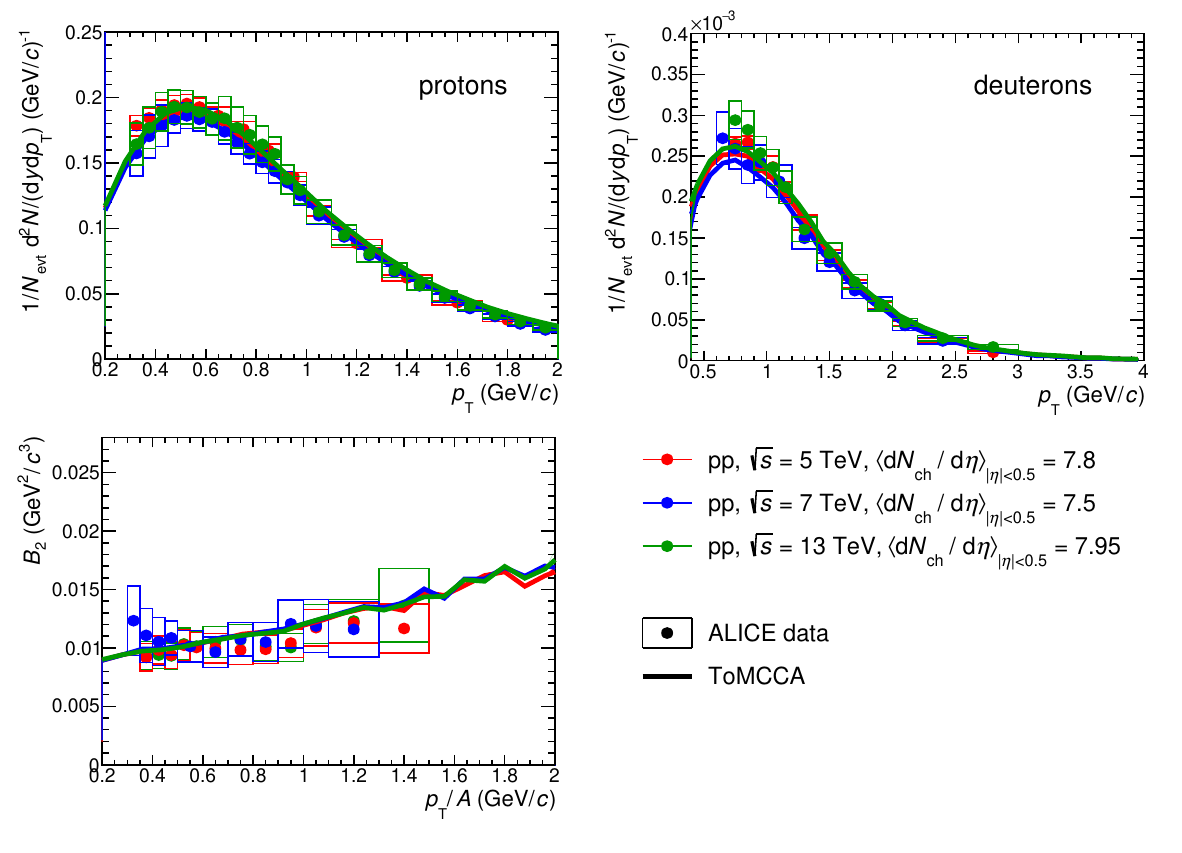}
    \caption{Results of ToMCCA in the multiplicity class VI, VII and VI+VII respectively for pp $\sqrt{s} =$ 5, 13 and 7 TeV, for the transverse momentum spectra of protons (top left) and deuterons (top right), and the coalescence parameter $B_2$ (bottom), compared to data measured by ALICE. The values of the average charged-particle multiplicity corresponding to each multiplicity class (VI, VII, and VI+VII) in the three different collision energies are indicated in the legend. }
    \label{fig:results_samemult}
\end{figure}

Using all the parameterizations as a function of multiplicity described in the previous Sections and the coalescence afterburner based on the Wigner function, it is possible to obtain the proton and deuteron momentum distributions for any value of multiplicity. In Fig.~\ref{fig:results_allclasses}, the resulting proton and deuteron momentum distributions obtained with ToMCCA are shown in the top panels and compared to the $p_{\mathrm{T}}$ distributions measured by ALICE~\cite{Acharya:2020sfy,13TeVHadrons,ALICE:2018pal,7TeVDeuterons,ALICE:2021ovi}. The different colors refer to different multiplicity classes, as reported in the legend, taken from different pp collision energies. All results show excellent agreement with the data points, being able to reproduce data within $1\sigma$ for all the multiplicity classes covered by pp collisions (up to $\sim36$). 
These results demonstrate that the coalescence model implemented in ToMCCA works for any collision energy and depends only on the chosen multiplicity class. This can be clearly seen in Fig.~\ref{fig:results_samemult}, where ToMCCA predictions are compared to results from three different energies of pp collisions, corresponding to similar multiplicity values ($dN_\mathrm{ch}/d\eta \sim 8$). Consequently, the model presented in this article can be used for any collision energy of pp collisions, finding applicability for the astrophysics searches of indirect dark matter. Indeed, to correctly interpret future measurements of the flux of antinuclei from cosmic rays, it is crucial to model the flux of antinuclei produced in collisions between the ISM and the primary CRs, which constitute the background for the DM searches. It has been shown in \cite{Serksnyte} that antideuterons in cosmic rays most likely originate from collisions of center-of-mass energy of $\sqrt{s}=20-30$ GeV.
The multiplicities corresponding to these collision energies have been measured by the ISR experiment~\cite{ISRMults} and correspond to values between $dN/d\eta=(1.50 \pm  0.01)$ and $dN/d\eta=(1.66\pm 0.02) $, for energies between $\sim 23.6\, \mathrm{GeV}$ and $\sim 30.8\, \mathrm{GeV}$. Hence, ToMCCA can be used to model the production of antideuterons at such multiplicities and to predict the expected flux of antinuclei due to the background.

Using the resulting proton and deuteron momentum distributions, the coalescence parameter $B_A$ can be computed. It is obtained from the ratio of the invariant yield of the nucleus with mass number $A$ and that of the protons squared, assuming protons and neutrons have the same momentum distributions as they are isospin partners, and $p_{\mathrm T}^{\mathrm p}$ = $p_{\mathrm T}^{A}$/$A$

\begin{equation}
B_{A} = { \biggl( \dfrac{1}{2 \pi p^{A}_{\mathrm T}} \biggl( \dfrac{ d^2N}{dydp_{\mathrm T}} \biggr)_{A}  \biggr)}  \bigg/{  \biggl( \dfrac{1}{2 \pi p^{p}_{\mathrm T}} \biggl( \dfrac{d^2N}{dydp_{\mathrm T}}  \biggr)_{p}  \biggr)^A}.  
\label{eq:BA}
\end{equation}

The resulting coalescence parameters are shown in the bottom panels of Figs.~\ref{fig:results_allclasses} and ~\ref{fig:results_samemult}, in comparison to ALICE measurements. Also in this case, the predictions of ToMCCA are in excellent agreement with the experimental results ($<1\sigma$). 

\subsection{Comparison with data as a function of multiplicity}
\begin{figure}
    \centering
    \resizebox{0.5\textwidth}{!}{
    \includegraphics{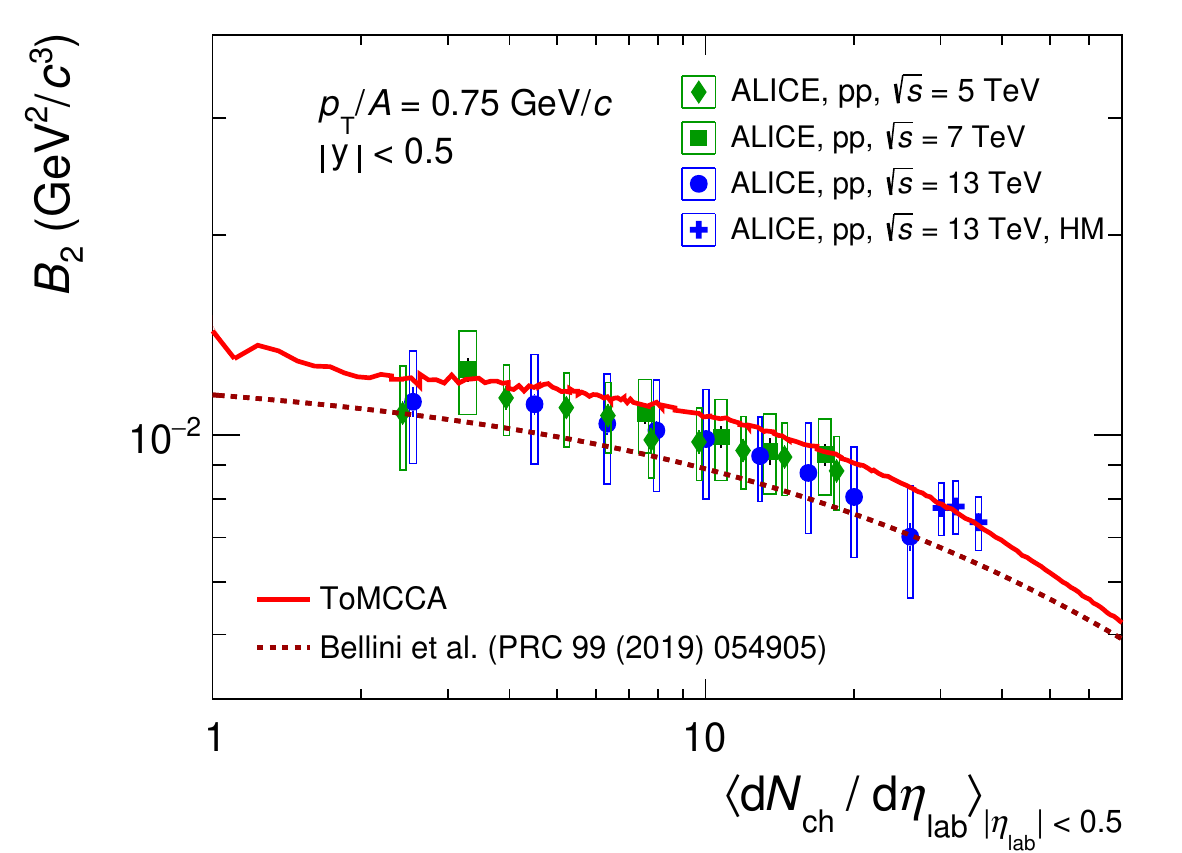}
    }
    \caption{Coalescence parameter $B_2$ as a function of multiplicity predicted by ToMCCA (red line), compared with the $B_2$ measurements from ALICE. The blue markers show the data included in the parametrizations, while the green markers show the ones not included. The dotted line is the model prediction of Ref.~\cite{Bellini:2018epz}, where the coalescence parameter for deuterons with a size of 3.2~fm is computed. }
    \label{fig:B2Mult}
\end{figure}

\begin{figure}
    \centering
    \resizebox{0.5\textwidth}{!}{
    \includegraphics{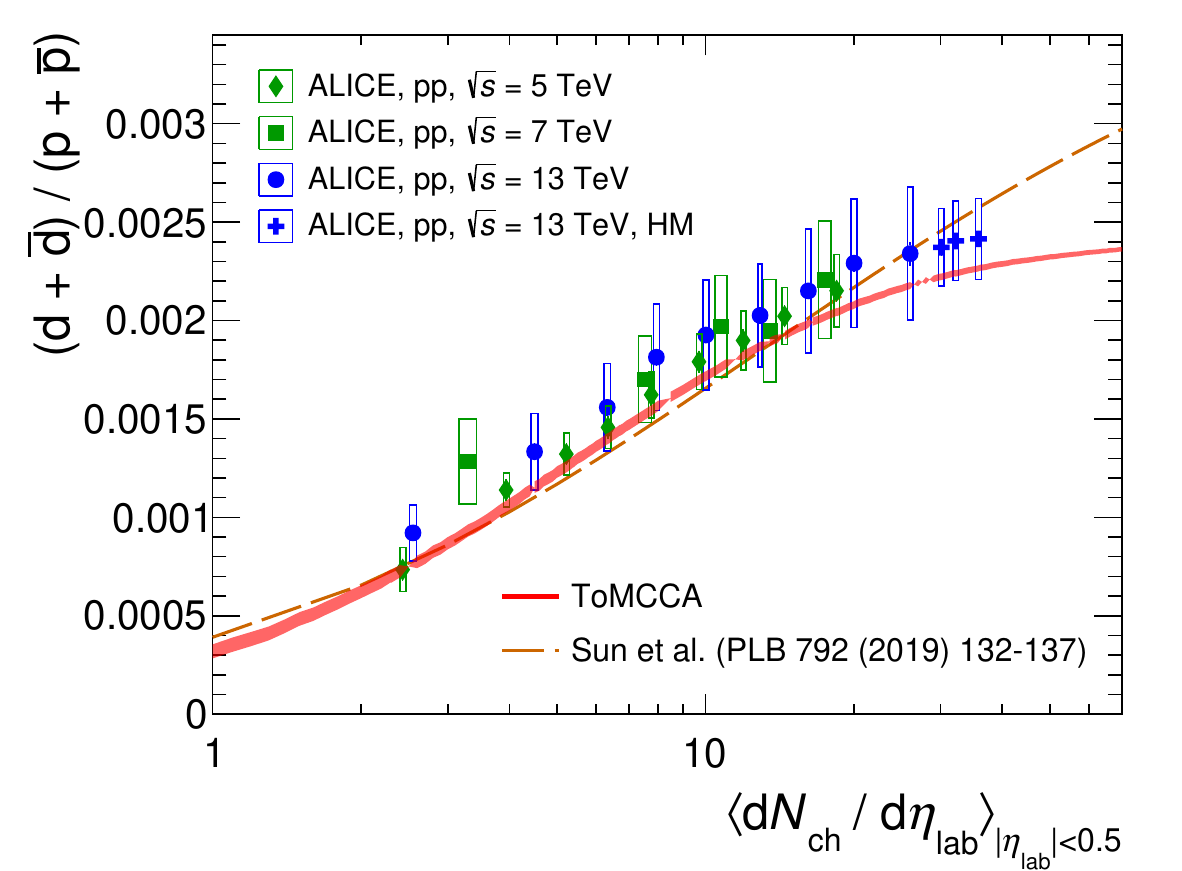}
    }
    \caption{Deuteron-to-proton ratio of integrated yields as a function of multiplicity obtained with ToMCCA (red solid line), compared to measurements of the ALICE Collaboration. The width of the model predictions reflects the statistical uncertainties. 
    The dotted line is the model prediction of Ref.~\cite{Coalescence6}. }
    \label{fig:dopMult}
\end{figure}

The results shown in the previous Sect.~\ref{results} can be obtained for arbitrary multiplicity. Hence, it is possible to scan the full range of average charged-particle multiplicity per event covered by pp collisions from $\sim1$ up to $\sim70$. Using the deuteron and proton $p_{\rm T}$ distributions obtained with ToMCCA, the coalescence parameter $B_2$ is computed, as well as the deuteron-to-proton ratio of the integrated yields (d/p). 
Selecting one value of $p_{\rm T}/A$, the predictions for the $B_2$ of ToMCCA are shown as a function of multiplicity in Fig.~\ref{fig:B2Mult}, and compared to the data measured by ALICE. Similarly, the d/p integrated yield ratios are shown as a function of multiplicity in Fig.~\ref{fig:dopMult}.
The predictions from ToMCCA show an excellent agreement with data, within 1$\sigma$, for both the $B_2$ and the d/p ratio. 

\subsection{Prediction of the Pearson correlation coefficient}
\begin{figure}
    \centering
    \resizebox{0.5\textwidth}{!}{
    \includegraphics{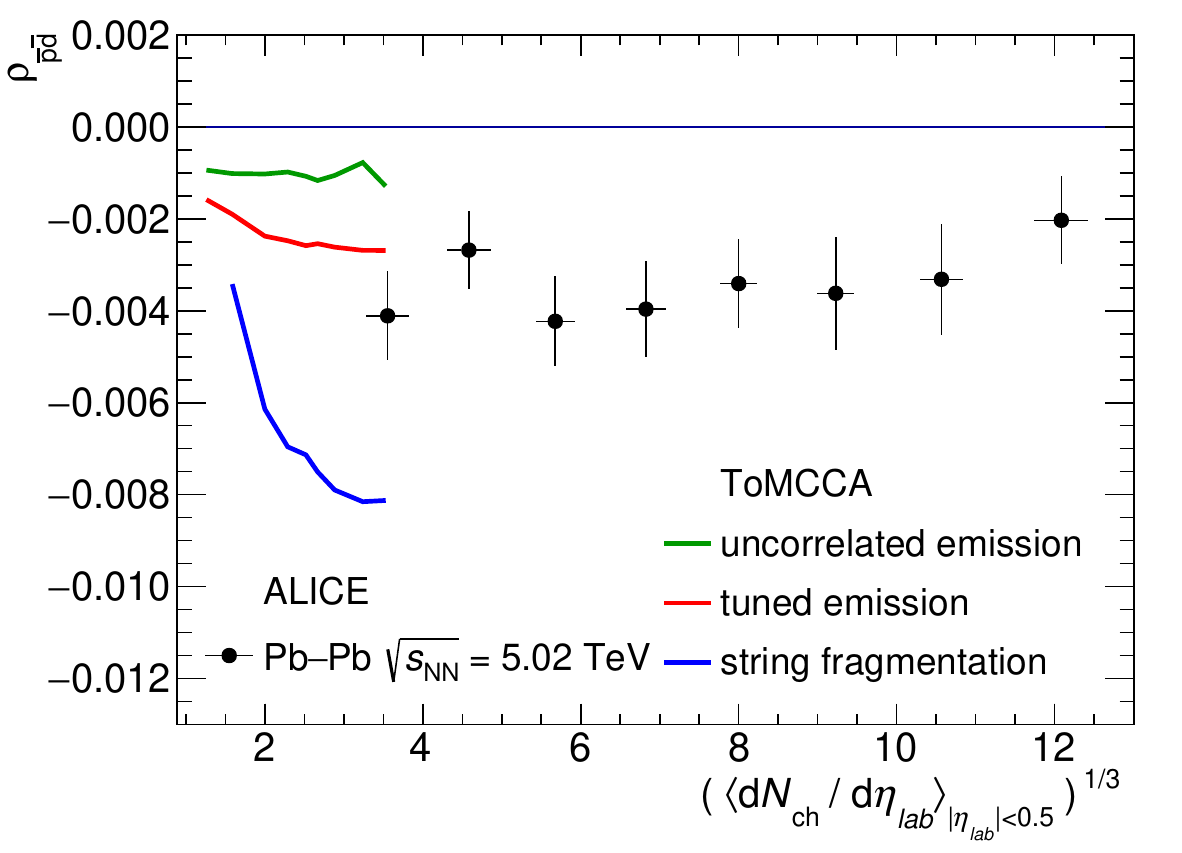}
    }
    \caption{Correlation coefficient for antiprotons and antideuterons ($\rho_{\overline{\rm p}\overline{\rm d}}$) as a function of the cubic root of the charged-particle multiplicity, predicted by ToMCCA, with the three different hadronization models, compared to the ALICE measurements in Pb--Pb collisions~\cite{FluctuationsPbPb}. }
    \label{fig:PearsonCoefficient}
\end{figure}

In Fig.~\ref{fig:PearsonCoefficient} the predicted Pearson correlation coefficient for antiprotons and antideuterons ($\rho_{\overline{\rm p}\overline{\rm d}}$), as defined in Eq.~\ref{eq:Pearson}, is shown. The predictions of ToMCCA are shown for the three hadronization models described in Sect.~\ref{sec:particleproduction}, i.e., Tuned Emission, String Fragmentation, and Uncorrelated Emission, within a multiplicity range of up to $N_{\rm{ch}}=44$. The source size, determined by the fit from Tuned Emission and used consistently throughout this paper, is used to calculate the $\rho_{\overline{\rm p}\overline{\rm d}}$ regardless of the hadronization model, in order to isolate the effect of baryon number suppression on the $\rho_{\overline{\rm p}\overline{\rm d}}$. These predictions are compared with measurements by the ALICE collaboration in Pb--Pb collisions at $\sqrt{s}_{\rm NN} =$ 5.02 TeV~\cite{FluctuationsPbPb}. The data points indicate that a non-zero $\rho_{\overline{\rm p}\overline{\rm d}}$ is necessary to match the observed suppression. However, it remains unclear if the transition from pp to Pb--Pb collisions is smooth. Future measurements of $\rho_{\overline{\rm p}\overline{\rm d}}$ in pp collisions could provide constraints on $\rho_{\overline{\rm p}\overline{\rm d}}$.

\section{Conclusions}

In this paper, we have introduced ToMCCA (Toy Monte Carlo Coalescence Afterburner), a toy Monte Carlo model for coalescence based on the Wigner function formalism. Our results show that ToMCCA is able to predict deuteron yields with a precision better than 5\% if all the required parameters, namely the charged particle multiplicity, nucleon momentum distributions, angular correlations, and the two-nucleon source size, are provided from parameterizations of experimental measurements. The results of this model demonstrate that ToMCCA can be used for arbitrary center-of-mass energy of pp collisions, as the coalescence predictions depend only on the average multiplicity of the event rather than on the collision energy. This would be useful for astrophysical studies, enabling the prediction of antinuclear fluxes in cosmic rays at the multiplicities corresponding to the energies at which antinuclei are produced in the collisions of ISM and CRs. This is key for disentangling the contribution of Dark Matter annihilations from the Standard Model background, for correctly interpreting future measurements of antinuclear fluxes in cosmic rays.

Due to its lightweight nature and minimum complexity, ToMCCA can also be easily extended to study more complex phenomena related to light nuclei formation, either towards other collision systems, lower collision energies and multiplicities, or towards heavier nuclei, as long as their wavefunction is known. 

\vspace{1cm}
\noindent

\section*{Declarations}

This work has received funding from the European Research Council (ERC) under the European Union's Horizon 2020 research and innovation programme (Grant Agreement No 950692).
This work has been supported by the Deutsche Forschungsgemeinschaft through grant SFB 1258 ``Neutrinos and Dark Matter in Astro- and Particle Physics''.
This research was supported by the Munich Institute for Astro- and Particle Physics (MIAPP) of the DFG cluster of excellence ``Origin and Structure of the Universe''.

\appendix
\section{Mathematical functions}
Power Law:

\begin{equation}
F(x) = A+B(x)^{C} .
\label{eq:PowerLaw}    
\end{equation}
Root-3-Sigmoid function, representing a smooth transition from cube-root behavior to zero-degree polynomial:

\begin{equation}
    F(x)=\frac{A}{1+e^{-B(x-M)}}x^{1/3}+(1-\frac{1}{1+e^{-B(x-M)}})C+D .
    \label{eq:Root3Sigmoid}
\end{equation}
Sigmoid Function:

\begin{equation}
    F(x)=\frac{A}{(B+e^{D\cdot x-C})}+N .
    \label{eq:Sigmoid}
\end{equation}
Levy-Tsallis, describing the transverse momentum distributions of particles with mass $m$, described by a temperature parameter $T$, a shape parameter $n$, and with total yield $dN/dy$: 

\begin{equation}
\dfrac {d^2 N}{dp_{\mathrm T} dy}=  \dfrac{dN}{dy} \dfrac{p_{\mathrm T}}{nT} \dfrac{(n-1) (n-2)}{(nT + m (n-2))} \left(1+ \dfrac {m_T - m}{nT} \right) ^{-n} .
\label{eq:LevyTsallis}
\end{equation}

%
\bibliographystyle{ieeetr}
\bibliography{Bibliography}

\end{document}